# Bayesian Vertex Nomination

May 23, 2012

Dominic S. Lee[1] and Carey E. Priebe[2]

[1] Department of Mathematics and Statistics, University of Canterbury
[2] Department of Applied Mathematics and Statistics, Johns Hopkins University

## Abstract

Consider an attributed graph $G$ with $n$ vertices, each of which is colored green or red, but only $m'$ vertices are observed to be red. The color of the other vertices is unobserved. Typically, the unknown total number, $m$, of red vertices is small, satisfying $m' \leq m \ll n$. The vertex nomination problem is to nominate one of the unobserved vertices as being red. The edge set of $G$ is a subset of the set of unordered pairs of vertices. Suppose that each edge is also colored green or red and this is observed for all edges. For a vertex, $v$, define its context statistic as the number of observed red vertices connected to $v$, and its content statistic as the number of red edges incident to $v$. Assuming that these statistics are independent between vertices and that red edges are more likely between red vertices, Coppersmith and Priebe (2012) proposed a likelihood model based on these statistics. Here, we formulate a Bayesian model using the proposed likelihood together with prior distributions chosen for the unknown parameters and unobserved vertex colors. From the resulting posterior distribution, the nominated vertex is the one with the highest posterior probability of being red. Inference with the model is conducted using a Metropolis-within-Gibbs algorithm, and performance is illustrated by a Monte Carlo simulation study. Simulation results show that (i) the Bayesian model performs significantly better than chance; (ii) there is evidence of a trend of increasing probability of correct nomination with increasing posterior probability that the nominated vertex is red; (iii) the Bayesian model performs increasingly better than the method in Coppersmith and Priebe, as the number of unobserved red vertices decreases relative to the total number of red vertices; and (iv) the rate of improvement in (iii) is higher when the total number of red vertices is bigger. An application example is provided using the Enron email corpus, where vertices represent Enron employees and their associates, observed red vertices are known fraudsters, red edges represent email communications perceived as fraudulent, and we wish to identify one of the latent vertices as most likely to be a fraudster.

# 1 Introduction

Suppose we have a community containing a small subset of interesting subjects. The identities of these interesting subjects are not fully known; in fact, only a few of them are known. The vertex nomination problem (Coppersmith and Priebe, 2012) is to nominate one of the unknown subjects as interesting and to do so with a quantifiable measure of being correct. This problem is distinct from but shares similarities with the Netflix challenge (Bell, et al., 2008), recommender systems (Resnick and Varian, 1997) and detecting communities of interest (Cortes, et al., 2002). Following Coppersmith and Priebe, our approach uses an attributed graph to model the community, with vertices representing subjects, a binary vertex attribute representing whether a subject is interesting, edges representing communications between subjects, and edge attributes representing contents of communications. By defining a context statistic (who communicates with who) and a content statistic (communication topics) associated with the attributed graph and making appropriate assumptions, Coppersmith and Priebe proposed a likelihood model for vertex nomination.

In this paper, we extend this likelihood model into a Bayesian model by introducing a vector of latent vertex attributes for the unknown subjects, together with appropriate prior distributions for parameters of the model. From the resulting posterior distribution, the nominated vertex is the one with the highest posterior probability of being interesting. Inference with the model is performed using a Metropolis-within-Gibbs algorithm. Performance of the model is illustrated using a Monte Carlo simulation study. Another simulation study compares its performance against the method in Coppersmith and Priebe. An application example is provided using the Enron email corpus (http://www.enron-mail.com/), where vertices represent Enron employees and their associates, edges represent email communications amongst them, and subjects of interest are those who had allegedly committed fraud. Other related inference problems had been investigated using this data set. Priebe, et al. (2005) looked at anomaly detection by using scan statistics on graphs derived from the email data. In Zhang, et al. (2006), an influence model that incorporated both email content information and context information (who emailed who) was used to learn the interaction matrix between people in the Enron corpus, and people were clustered based on this interaction matrix.

This paper is organized as follows. Details of the Bayesian model are described in the next section. The Markov chain Monte Carlo (MCMC) algorithm that implements the Bayesian solution is given in Section 3. Section 4 describes the simulation study and the results obtained. Experiments using the Enron email corpus are presented in Section 5. Finally, Section 6 summarizes and concludes the paper.

## 2 Model

We adopt the likelihood model proposed by Coppersmith and Priebe (2012) but with slightly different assumptions. Consider an attributed graph $G$ with $n$ vertices, each of which is colored green or red, but only $m'$ vertices ($1 < m' << n$) are observed to be red. The color of the other $n - m'$ vertices is unobserved or latent. Let the unknown total number of red vertices be $m$, satisfying $m' \leq m << n$. Thus, the number of latent green vertices is $n - m$ and the number of latent red vertices is $m - m'$; the latter can possibly be 0. Whilst Coppersmith and Priebe assumed that $0 < m' < m << n$, i.e. there is at least 1 observed red vertex and at least 1 latent red vertex, we assume that there are at least 2 observed red vertices but allow the number of latent red vertices to be 0. These new assumptions remain reasonable for our intended applications. More importantly, they allow simpler prior distributions that yield a simpler MCMC algorithm for implementing our Bayesian solution. If required, switching to Coppersmith and Priebe's original assumptions can be accommodated through a different choice of prior distribution that enforces the information that there is at least 1 latent red vertex. More details about this are given in the next section.

For a vertex $v$, let

(i) $Y(v)$ = vertex attribute of $v = \begin{cases} 1, & \text{if vertex is green,} \\ 2, & \text{if vertex is red;} \end{cases}$

(ii) $R(v)$ = number of observed red vertices connected to $v$;

(iii) $S(v)$ = number of red edges incident to $v$;

(iv) $T(v) = (R(v), S(v))$.

$R$ and $S$ are, respectively, the context and content statistics defined in Coppersmith and Priebe (2012). They showed that the use of both statistics resulted in better nomination performance than either one used alone. The advantage of using both context and content was also reported by Qi, et al. (2012), who modeled multimedia objects and their user-generated tags as a graph with context and content links for the purpose of multimedia annotation, which is analogous to vertex nomination where vertices are multimedia objects. For the Bayesian approach adopted here, there is potential to use the number of green edges incident to a vertex as an additional statistic, but at the cost of greater model and computational complexity. The cost-benefit of this added complexity is currently being investigated by the authors.

The edge set of *G* is a subset of the set of unordered pairs of vertices. The presence of an edge between two vertices denotes that the two vertices communicate. Each edge is also colored green or red, signifying the content of the communication, and this is observed for all edges. Note that *G* is an undirected graph with no self-loops, multi-edges or hyper-edges. For an edge *uv* between vertices *u* and *v*, let

(i) $Z(uv)$ = edge attribute of $uv = \begin{cases} 1, & \text{if edge is green,} \\ 2, & \text{if edge is red;} \end{cases}$

(ii) $p_1 = P(Z(uv) = 1 \mid Y(u) = Y(v) = 1 \text{ or } Y(u) \neq Y(v))$,
     = $P$(green edge between 2 green vertices or between a green vertex and a red one),
    $p_2 = P(Z(uv) = 2 \mid Y(u) = Y(v) = 1 \text{ or } Y(u) \neq Y(v))$,
     = $P$(red edge between 2 green vertices or between a green vertex and a red one),
    $p_0 = 1 - p_1 - p_2 = P(\text{no edge between } u \text{ and } v \mid Y(u) = Y(v) = 1 \text{ or } Y(u) \neq Y(v))$;

(iii) $q_1 = P(Z(uv) = 1 \mid Y(u) = Y(v) = 2)$,
     = $P$(green edge between 2 red vertices),
    $q_2 = P(Z(uv) = 2 \mid Y(u) = Y(v) = 2)$,
     = $P$(red edge between 2 red vertices),
    $q_0 = 1 - q_1 - q_2 = P(\text{no edge between } u \text{ and } v \mid Y(u) = Y(v) = 2)$.

The probabilities, $q_0$, $q_1$ and $q_2$, quantify both the frequency $(q_1 + q_2)$ of communication and distribution $(q_1, q_2)$ of content amongst red vertices. Likewise, $p_0$, $p_1$ and $p_2$ quantify these for the rest of the graph, i.e. amongst green vertices as well as between a red vertex and a green one. Two key assumptions underpinning Coppersmith and Priebe's model are (i) pairs of red vertices, both observed and latent, communicate with a different frequency from other pairs; and (ii) the distribution of content amongst red vertices is different from the rest of the graph. More specifically, it is assumed that $p_1 = q_1$ and $p_2 < q_2$, where the latter prescribes that red edges are more likely between red vertices, and hence a higher frequency of communication amongst red vertices $(p_1 + p_2 < q_1 + q_2)$.

Assuming that the context and content statistics are independent between vertices, their joint distribution can be described as follows. Given that *v* is a green vertex,

$$f_1(R(v) \mid p_1, p_2) = Bin(m', p_1 + p_2), \tag{1}$$

$$f_1(S(v) \mid p_2) = Bin(n - 1, p_2), \tag{2}$$

$$f_1(T(v) | p_1, p_2) = f_1(S(v) | R(v), p_1, p_2) f_1(R(v) | p_1, p_2)$$
$$= \{Bin(n - m' - 1, p_2) * Bin(R(v), \tfrac{p_2}{p_1 + p_2})\} \cdot Bin(m', p_1 + p_2). \quad (3)$$

Here, $Bin(n, p)$ represents a binomial mass function with parameters $n$ and $p$, $g*h$ denotes the discrete convolution,

$$g * h(y) = \sum_z g(y - z) h(z), \quad (4)$$

and $f * g * h$ is the double convolution,

$$f * g * h(x) = \sum_y f(x - y) \sum_z g(y - z) h(z). \quad (5)$$

Given that $v$ is a latent red vertex,

$$f_2(R(v) | p_1, q_2) = Bin(m', p_1 + q_2), \quad (6)$$

$$f_2(S(v) | m, p_2, q_2) = Bin(n - m, p_2) * Bin(m - 1, q_2), \quad (7)$$

$$f_2(T(v) | m, p_1, p_2, q_2)$$
$$= f_2(S(v) | R(v), m, p_1, p_2, q_2) f_2(R(v) | p_1, q_2) \quad (8)$$
$$= \{Bin(n - m, p_2) * Bin(m - m' - 1, q_2) * Bin(R(v), \tfrac{q_2}{p_1 + q_2})\} \cdot Bin(m', p_1 + q_2).$$

Given that $v$ is an observed red vertex,

$$f'(R(v) | p_1, q_2) = Bin(m' - 1, p_1 + q_2), \quad (9)$$

$$f'(S(v) | m, p_2, q_2) = Bin(n - m, p_2) * Bin(m - 1, q_2), \quad (10)$$

$$f'(T(v) | m, p_1, p_2, q_2)$$
$$= f'(S(v) | R(v), m, p_1, p_2, q_2) f'(R(v) | p_1, q_2) \quad (11)$$
$$= \{Bin(n - m, p_2) * Bin(m - m', q_2) * Bin(R(v), \tfrac{q_2}{p_1 + q_2})\} \cdot Bin(m' - 1, p_1 + q_2).$$

Let $\mathbf{T'} = \{T'(1), ..., T'(m')\}$ be the statistics for those vertices whose attributes are observed to be red. Similarly, let $\mathbf{T} = \{T(1), ..., T(n - m')\}$ be the statistics for those vertices whose attributes, $\mathbf{Y} = \{Y(1), ..., Y(n - m')\}$, are unknown. By making the simplifying assumption that the statistics are conditionally independent given **Y**, $p_1$, $p_2$ and $q_2$, the likelihood function is given by

$$f(\mathbf{T},\mathbf{T}' \mid \mathbf{Y}, p_1, p_2, q_2)$$
$$= \prod_{i:Y(i)=1} f_1(T(i) \mid p_1, p_2) \prod_{j:Y(j)=2} f_2(T(j) \mid m, p_1, p_2, q_2) \prod_{k=1}^{m'} f'(T'(k) \mid m, p_1, p_2, q_2), \quad (12)$$

where $m = m' + \sum_{i=1}^{n-m'} I_{\{2\}}(Y(i))$.

By Bayes rule, the posterior distribution for the unknown quantities, **Y**, $p_1$, $p_2$ and $q_2$, is given by

$$f(\mathbf{Y}, p_1, p_2, q_2 \mid \mathbf{T},\mathbf{T}') \propto f(\mathbf{T},\mathbf{T}' \mid \mathbf{Y}, p_1, p_2, q_2) f(\mathbf{Y}, p_1 p_2, q_2), \quad (13)$$

where $f(\mathbf{Y}, p_1, p_2, q_2)$ is a prior distribution that must be specified. For our problem, the latent attribute vector, **Y**, is the quantity of interest while $p_1$, $p_2$ and $q_2$ may be regarded as nuisance parameters.

We assume, for the prior distribution, that **Y** is independent of $(p_1, p_2, q_2)$, and choose conditionally independent Bernoulli($\psi$) distributions for the components of **Y**, where $\psi = P(Y_i = 2)$ requires a hyperprior distribution. For $(p_1, p_2, q_2)$, we choose a Dirichlet distribution for $(p_1, p_2)$ and a uniform distribution for $q_2$ conditional on $p_1$ and $p_2$. We thus have

$$f(\mathbf{Y}, p_1, p_2, q_2 \mid \psi) = f(\mathbf{Y} \mid \psi) f(p_1, p_2, q_2), \quad (14)$$

with

$$f(\mathbf{Y} \mid \psi) = \prod_{i=1}^{n-m'} Bernoulli(\psi) = \psi^{m-m'}(1-\psi)^{n-m}, \quad (15)$$

and

$$f(p_1, p_2, q_2) = f(q_2 \mid p_1, p_2) f(p_1, p_2)$$
$$= U(p_2, 1-p_1) \cdot Dir(\alpha_0, \alpha_1, \alpha_2). \quad (16)$$

Choosing $\alpha_0 = \alpha_1 = \alpha_2 = 1$ for the Dirichlet distribution gives

$$f(p_1, p_2, q_2) = \frac{2}{1 - p_1 - p_2} I_{(0,1)}(p_1) I_{(0,1-p_1)}(p_2) I_{(p_2, 1-p_1)}(q_2), \tag{17}$$

with marginal prior distributions,

$$f(p_1) = f(p_2) = beta(1,2), \tag{18}$$

$$f(q_2) = 2\left[ q_2 \log\left(\frac{1-q_2}{q_2}\right) - \log(1-q_2) \right] I_{(0,1)}(q_2). \tag{19}$$

An obvious choice of hyperprior distribution for $\psi$ is the beta distribution with parameters $\alpha$, $\beta > 0$:

$$f(\psi \mid \alpha, \beta) \propto \psi^{\alpha-1}(1-\psi)^{\beta-1} I_{(0,1)}(\psi). \tag{20}$$

The posterior distribution can now be written as

$$\begin{aligned} f(\mathbf{Y}, p_1, p_2, q_2, \psi \mid \mathbf{T}, \mathbf{T}') \propto{} & f(\mathbf{T}, \mathbf{T}' \mid \mathbf{Y}, p_1, p_2, q_2) \psi^{m-m'+\alpha-1}(1-\psi)^{n-m+\beta-1} I_{(0,1)}(\psi) \\ & \cdot (1 - p_1 - p_2)^{-1} I_{(0,1)}(p_1) I_{(0,1-p_1)}(p_2) I_{(p_2, 1-p_1)}(q_2). \end{aligned} \tag{21}$$

## 3 Inference

Posterior inference can proceed via MCMC using a Metropolis within Gibbs algorithm. Since the components of **Y** are binary, they can be updated sequentially using Gibbs sampling as follows. Let $\mathbf{Y_{-i}} = \mathbf{Y} \setminus Y(i)$ and let $\gamma_i$ be the conditional posterior probability that the unlabelled vertex $i$ is red given the attributes $\mathbf{Y_{-i}}$. Thus,

$$\begin{aligned} &\gamma_i(\mathbf{Y}_{-i}, p_1, p_2, q_2, \psi) \\ &= P(Y(i) = 2 \mid \mathbf{Y}_{-i}, \mathbf{T}, \mathbf{T}', p_1, p_2, q_2, \psi) \\ &= \frac{f(Y(i) = 2, \mathbf{Y}_{-\mathbf{i}}, p_1, p_2, q_2, \psi \mid \mathbf{T}, \mathbf{T}')}{f(Y(i) = 1, \mathbf{Y}_{-\mathbf{i}}, p_1, p_2, q_2, \psi \mid \mathbf{T}, \mathbf{T}') + f(Y(i) = 2, \mathbf{Y}_{-\mathbf{i}}, p_1, p_2, q_2, \psi \mid \mathbf{T}, \mathbf{T}')}, \end{aligned} \tag{22}$$

or

$$\frac{1}{\gamma_i(\mathbf{Y_{-i}},p_1,p_2,q_2,\psi)}$$

$$= 1 + \frac{f(Y(i)=1,\mathbf{Y_{-i}},p_1,p_2,q_2,\psi \mid \mathbf{T},\mathbf{T'})}{f(Y(i)=2,\mathbf{Y_{-i}},p_1,p_2,q_2,\psi \mid \mathbf{T},\mathbf{T'})}$$

$$= 1 + \frac{[(1-\psi)/\psi]\, f_1(T(i) \mid p_1,p_2) \prod_{j:j\neq i,Y(j)=2} f_2(T(j) \mid m_{-i},p_1,p_2,q_2)\prod_{k=1}^{m'} f'(T'(k) \mid m_{-i},p_1,p_2,q_2)}{f_2(T(i) \mid m_{-i}+1,p_1,p_2,q_2) \prod_{j:j\neq i,Y(j)=2} f_2(T(j) \mid m_{-i}+1,p_1,p_2,q_2)\prod_{k=1}^{m'} f'(T'(k) \mid m_{-i}+1,p_1,p_2,q_2)}$$

$$= 1 + \frac{[(1-\psi)/\psi]\, f_1(T(i) \mid p_1,p_2) \prod_{j:j\neq i,Y(j)=2} f_2(S(j) \mid R(j),m_{-i},p_1,p_2,q_2)\prod_{k=1}^{m'} f'(S'(k) \mid R'(k),m_{-i},p_1,p_2,q_2)}{f_2(T(i) \mid m_{-i}+1,p_1,p_2,q_2) \prod_{j:j\neq i,Y(j)=2} f_2(S(j) \mid R(j),m_{-i}+1,p_1,p_2,q_2)\prod_{k=1}^{m'} f'(S'(k) \mid R'(k),m_{-i}+1,p_1,p_2,q_2)},$$

(23)

where $m_{-i} = m' + \sum_{j:j\neq i} I_{\{2\}}(Y(j))$.

The conditional posterior density of $\psi$ given $\mathbf{Y}$ and $(p_1,p_2,q_2)$ is

$$f(\psi \mid \mathbf{T},\mathbf{T'},\mathbf{Y},p_1,p_2,q_2) \propto \psi^{m-m'+\alpha-1}(1-\psi)^{n-m+\beta-1} I_{(0,1)}(\psi), \tag{24}$$

which is the $beta(m-m'+\alpha, n-m+\beta)$ density. Thus, $\psi$ can easily be updated using a Gibbs update step. Unfortunately, the conditional posterior distribution of $(p_1,p_2,q_2)$ given $\mathbf{Y}$ and $\psi$ does not have a standard form that we can generate from exactly. As such, we use random-walk Metropolis-Hastings to update each parameter in turn, using the conditional distributions, $f(p_1 \mid p_2,q_2)$, $f(p_2 \mid p_1,q_2)$ and $f(q_2 \mid p_1,p_2)$, obtained from (17) as proposal distributions. The last of these is, by choice,

$$f(q_2 \mid p_1,p_2) = U(p_2, 1-p_1) = (1-p_1-p_2)^{-1} I_{(p_2,1-p_1)}(q_2). \tag{25}$$

From (17), the conditional prior for $p_1$ given $p_2$ and $q_2$ is

$$f(p_1 \mid p_2,q_2) \propto (1-p_1-p_2)^{-1} I_{(0,1-q_2)}(p_1), \tag{26}$$

which, for $p_1 \in (0, 1-q_2)$, can be written as

$$f(p_1 \mid p_2,q_2) = \frac{(1-p_1-p_2)^{-1}}{\log(1-p_2)-\log(q_2-p_2)}. \tag{27}$$

The corresponding conditional distribution function is

$$F(p_1 \mid p_2, q_2) = \frac{\log(1-p_2) - \log(1-p_1-p_2)}{\log(1-p_2) - \log(q_2-p_2)}, \tag{28}$$

and so the conditional inverse distribution function is, for $u \in [0,1]$,

$$F^{-1}(u \mid p_2, q_2) = 1 - p_2 - \frac{(q_2-p_2)^u}{(1-p_2)^{u-1}}. \tag{29}$$

This enables us to generate from $f(p_1 \mid p_2, q_2)$ easily by the inverse distribution function method. Similarly,

$$f(p_2 \mid p_1, q_2) \propto (1-p_1-p_2)^{-1} I_{(0,q_2)}(p_2), \tag{30}$$

and so for $p_2 \in (0, q_2)$,

$$f(p_2 \mid p_1, q_2) = \frac{(1-p_1-p_2)^{-1}}{\log(1-p_1) - \log(1-p_1-q_2)}, \tag{31}$$

$$F(p_2 \mid p_1, q_2) = \frac{\log(1-p_1) - \log(1-p_1-p_2)}{\log(1-p_1) - \log(1-p_1-q_2)}, \tag{32}$$

and

$$F^{-1}(u \mid p_1, q_2) = 1 - p_1 - \frac{(1-p_1-q_2)^u}{(1-p_1)^{u-1}}. \tag{33}$$

Finally, denoting the state at iteration $h$ by $(\mathbf{Y}^{(h)}, p_1^{(h)}, p_2^{(h)}, q_2^{(h)}, \psi^{(h)})$, our Metropolis-within-Gibbs sampler proceeds according to:

Gibbs step:

For $i = 1, \ldots, n - m'$:

Compute $\gamma_i(Y^{(h)}(1), \ldots, Y^{(h)}(i-1), Y^{(h-1)}(i+1), Y^{(h-1)}(n-m'), p_1^{(h-1)}, p_2^{(h-1)}, q_2^{(h-1)}, \psi^{(h-1)})$,

$$\text{Set } Y^{(h)}(i) = \begin{cases} 1 & \text{with probability } 1 - \gamma_i, \\ 2 & \text{with probability } \gamma_i. \end{cases}$$

$$\text{Compute } m^{(h)} = m' + \sum_{i=1}^{n-m'} I_{\{2\}}(Y^{(h)}(i)),$$

$$\text{Generate } \psi^{(h)} \sim beta(m^{(h)} - m' + \alpha, n - m^{(h)} + \beta).$$

Metropolis-Hastings step:

$$\text{Generate } p_1^* \sim f(p_1 \mid p_2^{(h-1)}, q_2^{(h-1)}),$$

$$\text{Compute } \pi(p_1) = \min\left\{1, \frac{f(\mathbf{T}, \mathbf{T}' \mid \mathbf{Y}^{(h)}, p_1^*, p_2^{(h-1)}, q_2^{(h-1)})}{f(\mathbf{T}, \mathbf{T}' \mid \mathbf{Y}^{(h)}, p_1^{(h-1)}, p_2^{(h-1)}, q_2^{(h-1)})}\right\},$$

$$\text{Set } p_1^{(h)} = \begin{cases} p_1^* & \text{with probability } \pi(p_1), \\ p_1^{(h-1)} & \text{with probability } 1 - \pi(p_1), \end{cases}$$

$$\text{Generate } p_2^* \sim f(p_2 \mid p_1^{(h)}, q_2^{(h-1)}),$$

$$\text{Compute } \pi(p_2) = \min\left\{1, \frac{f(\mathbf{T}, \mathbf{T}' \mid \mathbf{Y}^{(h)}, p_1^{(h)}, p_2^*, q_2^{(h-1)})}{f(\mathbf{T}, \mathbf{T}' \mid \mathbf{Y}^{(h)}, p_1^{(h)}, p_2^{(h-1)}, q_2^{(h-1)})}\right\},$$

$$\text{Set } p_2^{(h)} = \begin{cases} p_2^* & \text{with probability } \pi(p_2), \\ p_2^{(h-1)} & \text{with probability } 1 - \pi(p_2), \end{cases}$$

$$\text{Generate } q_2^* \sim f(q_2 \mid p_1^{(h)}, p_2^{(h)}),$$

$$\text{Compute } \pi(q_2) = \min\left\{1, \frac{f(\mathbf{T}, \mathbf{T}' \mid \mathbf{Y}^{(h)}, p_1^{(h)}, p_2^{(h)}, q_2^*)}{f(\mathbf{T}, \mathbf{T}' \mid \mathbf{Y}^{(h)}, p_1^{(h)}, p_2^{(h)}, q_2^{(h-1)})}\right\},$$

$$\text{Set } q_2^{(h)} = \begin{cases} q_2^* & \text{with probability } \pi(q_2), \\ q_2^{(h-1)} & \text{with probability } 1 - \pi(q_2). \end{cases}$$

The ability to update $\psi$ using a Gibbs step that generates from a beta distribution is the motivation for allowing the number of latent red vertices to be 0. This makes the independent Bernoulli model in (15) a possible choice as prior for $\mathbf{Y}$. Together with the beta hyperprior for $\psi$ in (20), we end up with a conjugate conditional posterior in (24) that facilitates the Gibbs step for updating $\psi$. With Coppersmith and Priebe's (2012) assumption that the number of latent red vertices is at least 1, however, the prior in (15) is no longer appropriate. Letting $\Sigma_{\mathbf{Y}} = Y(1) + \cdots + Y(n - m')$, a possible alternative is

$$f(\mathbf{Y} \mid \psi) = \begin{cases} 0, & \Sigma_{\mathbf{Y}} = 0, \\ \dfrac{\psi^{m-m'}(1-\psi)^{n-m}}{1-(1-\psi)^{n-m'}}, & \Sigma_{\mathbf{Y}} > 0, \end{cases} \quad (34)$$

which no longer admits a conjugate hyperprior. Updating of $\psi$ will therefore require an additional Metropolis-Hastings step within the Gibbs sampler. The cost-benefit of using this alternative prior model is currently being investigated by the authors.

## 4  Simulation Results

Consider an illustrative example where $n = 12$, $m = 5$, $m' = 2$, $p_1 = 0.25$, $p_2 = 0.15$ and $q_2 = 0.25$. A particular graph realization is shown in Figure 1. Labeling the observed red vertices as 1 and 2, the latent red vertices as 3, 4 and 5, and the latent green vertices from 6 to 12, the observed edge attributes are given in Table 1.

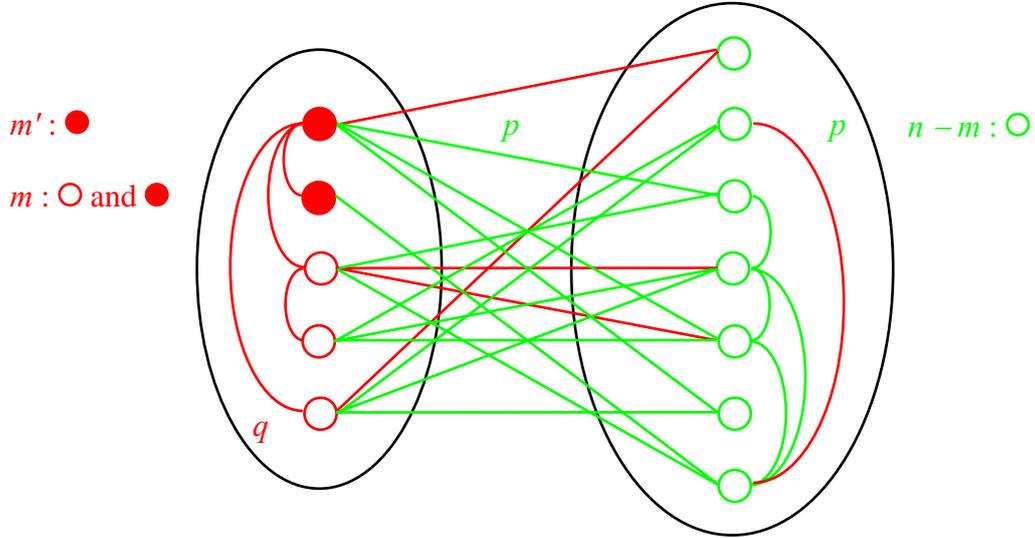

Figure 1. Illustrative attributed graph with 12 vertices corresponding to the adjacency matrix in Table 1. Here, $m' = 2$ vertices are observed to be red, $m - m' = 3$ are latent red vertices and $n - m = 7$ are latent green vertices. Edges represent communication between connected vertices and edge attributes, denoting content of communication, are assumed to be binary: green or red (1 or 2, respectively, in Table 1). The frequency of communication and distribution of content amongst red vertices are governed by $q = (q_0, q_1, q_2)$, whilst $p = (p_0, p_1, p_2)$ quantifies these for the rest of the graph, i.e. amongst green vertices as well as between a red vertex and a green one. Assuming that all edges and their attributes are observed, these are used together with the observed red vertices to nominate one of the latent vertices as being red.

Table 1. Binary edge attributes (1 = green, 2 = red) for the attributed graph shown in Figure 1. An entry of 0 denotes the absence of an edge.

| | | ● | ○ | ○ | ○ | ○ | ○ | ○ | ○ | ○ | ○ | ○ |
|---|---|---|---|---|---|---|---|---|---|---|---|---|
| | | 2 | 3 | 4 | 5 | 6 | 7 | 8 | 9 | 10 | 11 | 12 |
| ● | 1 | 2 | 2 | 0 | 2 | 2 | 0 | 1 | 0 | 1 | 1 | 0 |
| ● | 2 | | 0 | 0 | 0 | 0 | 0 | 0 | 0 | 0 | 0 | 1 |
| ○ | 3 | | | 2 | 0 | 0 | 0 | 1 | 2 | 2 | 0 | 1 |
| ○ | 4 | | | | 0 | 0 | 1 | 0 | 1 | 1 | 0 | 0 |
| ○ | 5 | | | | | 2 | 1 | 0 | 1 | 0 | 1 | 0 |
| ○ | 6 | | | | | | 0 | 0 | 0 | 0 | 0 | 0 |
| ○ | 7 | | | | | | | 0 | 0 | 0 | 0 | 2 |
| ○ | 8 | | | | | | | | 1 | 0 | 0 | 0 |
| ○ | 9 | | | | | | | | | 1 | 0 | 1 |
| ○ | 10 | | | | | | | | | | 0 | 1 |
| ○ | 11 | | | | | | | | | | | 0 |

To use the MCMC sampler developed, values must first be specified for the parameters of the beta hyperprior for $\psi$. Our choice is motivated by the overall goal to nominate a *single* vertex as being red. Hence, it is desirable that the hyperprior be chosen to induce sparsity in the potential nominees. One way to achieve this is to select a beta density with mode at $1/(n-m')$; a convenient choice being $\alpha = 2$ and $\beta = n - m'$. The results given in this section and the next are based on this choice. We have also studied other choices, including (i) a flat prior (i.e. $beta(1,1)$ or $U(0,1)$); (ii) a flat prior on the interval (0, 0.5) (i.e. truncated $beta(1,1)$ on (0, 0.5) or $U(0, 0.5)$) motivated by our expectation that $n \gg m$; and (iii) a beta density with $\alpha = m'$ and $\beta = n - 2m'$, and thus having mean at $m'/(n-m')$. Results for these other choices are not reported here but we observed that inference about the probability of correct nomination and nuisance parameters was insensitive to the choice of hyperprior even though there were variations in posterior inference about latent vertex attributes and the hyperparameter $\psi$.

20000 MCMC iterations were performed for this graph and trace plots for the latent attributes, nuisance parameters and hyperparameter are given in Figures 2, 3 and 4. Figure 2 shows trace plots of the moving average estimates of the marginal posterior probabilities that each of the 10 unlabelled vertices is red. For unlabelled vertex $i$, for example, this is estimated at iteration $h$ by

$$\hat{P}_h(Y(i) = 2 \mid \mathbf{T}, \mathbf{T}') = \frac{1}{h} \sum_{j=1}^{h} I_{\{2\}}(Y^{(j)}(i)). \tag{35}$$

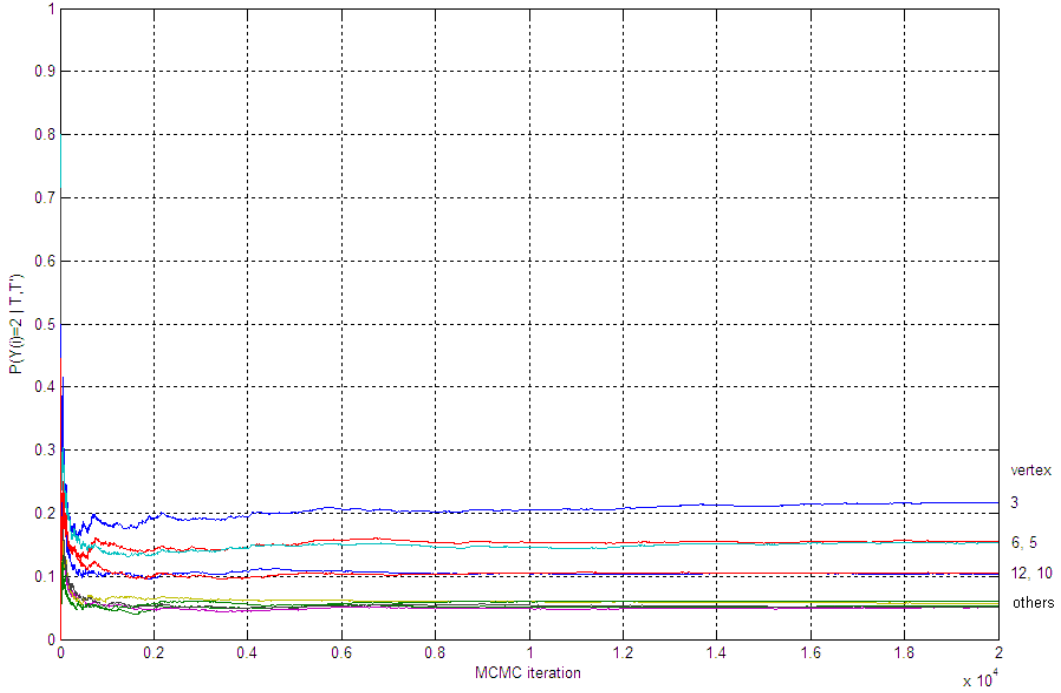

Figure 2. Trace plots of the moving average estimates of the marginal posterior probabilities that each of the unlabelled vertices is red. The top-five ranking vertices (3, 6, 5, 12, 10) are labeled as shown; the others (11, 4, 8, 7, 9) are clustered together at the bottom. Recall that the three latent red vertices are 3, 4 and 5, and so we have a correct nomination in this case.

Figures 3 shows trace plots of the nuisance parameters and hyperparameter, while Figure 4 shows moving average estimates of their marginal posterior means. For example, for $p_1$ in the first plot in Figure 4, its marginal posterior mean at iteration $h$ is estimated by

$$\hat{E}_h(p_1 \mid \mathbf{T}, \mathbf{T}') = \frac{1}{h} \sum_{j=1}^{h} p_1^{(j)}. \tag{36}$$

Based on Figures 2 and 3, and primarily on Figure 2 since these probabilities are the quantities of interest in our problem, we discarded the first 10000 iterations as burn-in and used the last 10000 for posterior inference. Using the last 10000 iterations, the sample autocorrelations for the first unlabelled vertex (vertex number 3) up to lag 100 are shown in Figure 5, where rapid decay in correlation is evident. The sample autocorrelations for the other unlabelled vertices are not shown but they all exhibit similar rapid decay.

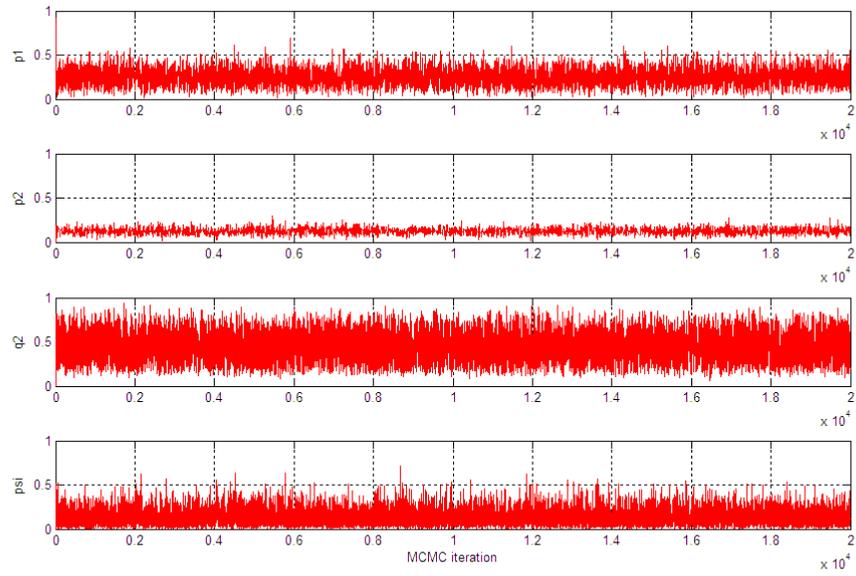

Figure 3. Trace plots of nuisance parameters, $p_1$, $p_2$, $q_2$, and hyperparameter, $\psi$.

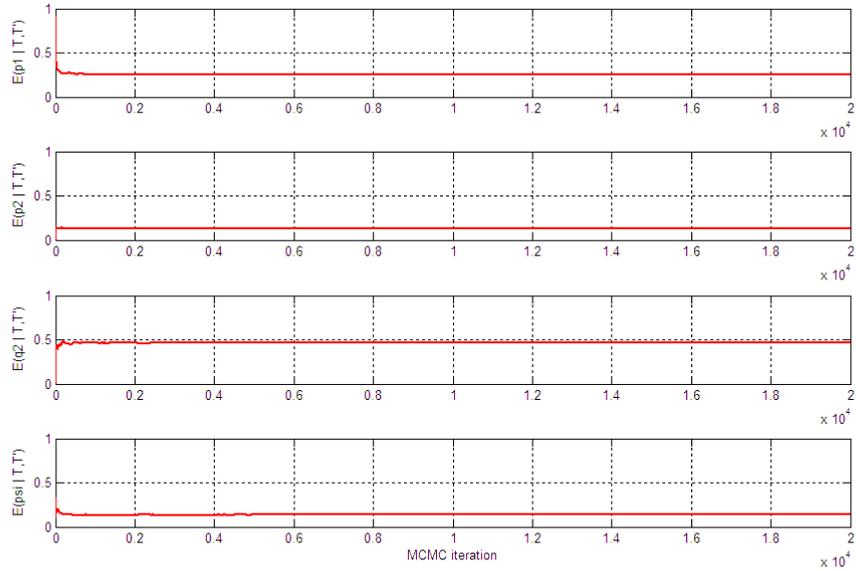

Figure 4. Trace plots of moving average estimates of the marginal posterior means of the nuisance parameters, $p_1$, $p_2$, $q_2$, and hyperparameter, $\psi$.

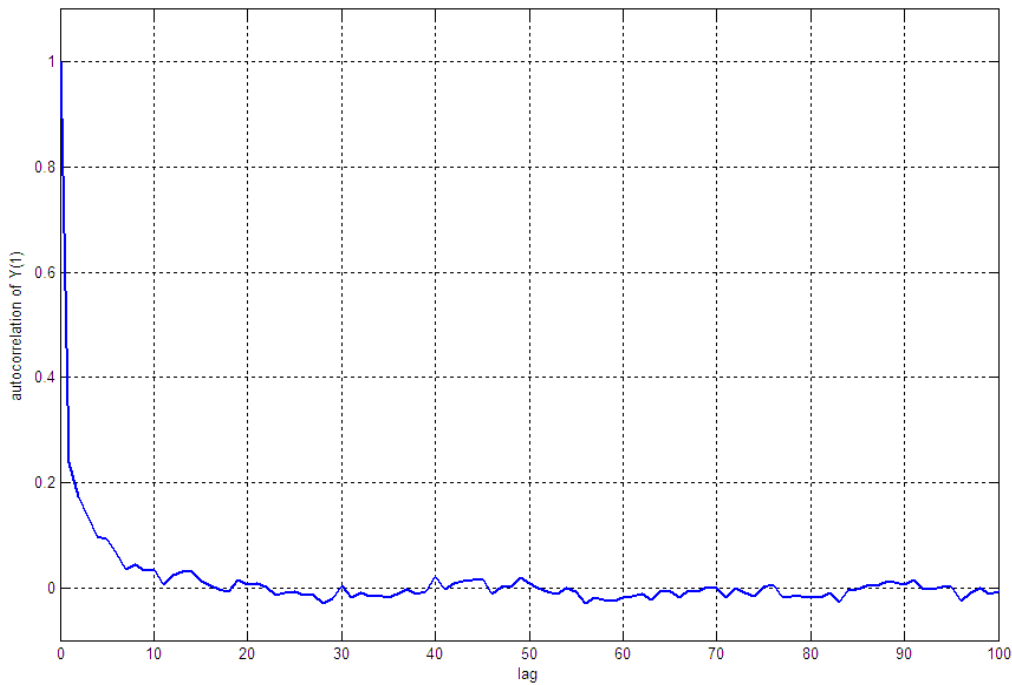

Figure 5. Sample autocorrelations up to lag 100 for the first unlabelled vertex (i.e. vertex number 3).

Based on the last 10000 MCMC iterations, estimates of the marginal posterior probabilities that each of the unlabelled vertices is a red vertex are given in Table 2. Consequently, vertex number 3, which has the maximum probability, will be nominated. In this case, this turns out to be a correct nomination (recall that the latent red vertices are 3, 4 and 5). We advise caution in interpreting these marginal posterior probabilities at face values because the relationship between them and the probability of correct nomination is not so straightforward. Even though we observed that the rankings of these posterior probabilities were quite insensitive to the hyperprior for $\psi$, the values of the posterior probabilities do vary with different hyperpriors. Fortunately, we will show later on that there is evidence of a trend of increasing probability of correct nomination with increasing maximum marginal posterior probability of a latent vertex being red.

Although inference about the nuisance parameters and hyperparameter is not required, it is interesting to look at their prior and posterior distributions. The marginal prior and posterior densities are shown in Figure 6, using kernel density estimates constructed from the last 10000 MCMC iterations. We used a Gaussian kernel density estimator with diffusion-based bandwidth selection (Algorithm 1 in Botev, et al., 2010).

Table 2. Posterior probabilities that latent vertex is red for the illustrative attributed graph with 12 vertices.

| Vertex number | $\hat{P}(Y(i) = 2 \mid \mathbf{T}, \mathbf{T}')$ |
| --- | --- |
| 3 | 0.2281 |
| 4 | 0.0550 |
| 5 | 0.1551 |
| 6 | 0.1596 |
| 7 | 0.0519 |
| 8 | 0.0543 |
| 9 | 0.0496 |
| 10 | 0.1031 |
| 11 | 0.0603 |
| 12 | 0.1045 |

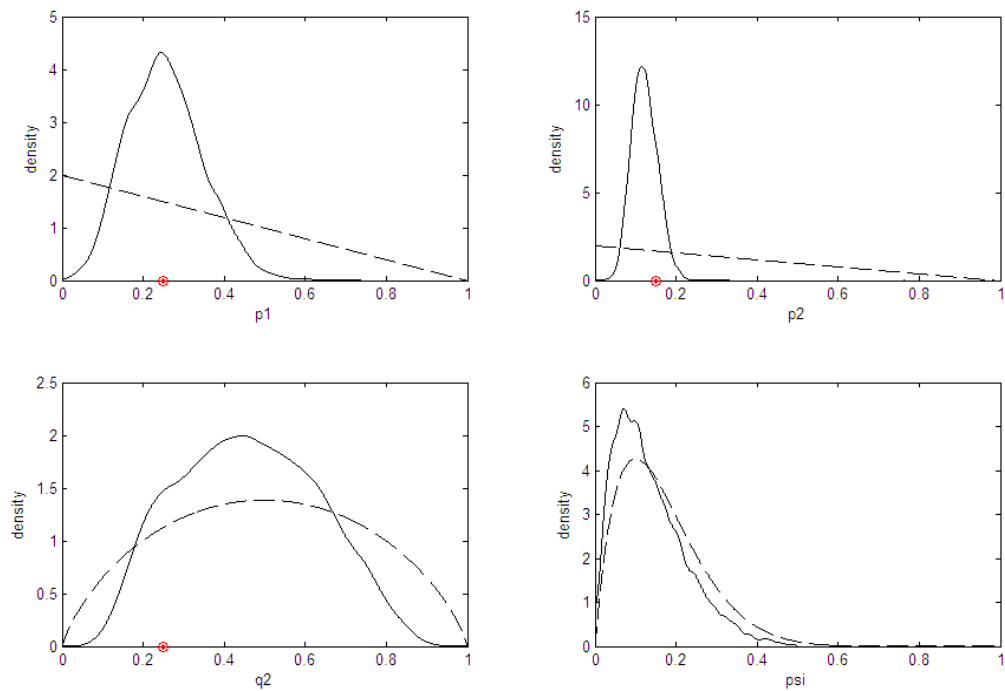

Figure 6. Marginal prior densities (dashed curves) and posterior densities (solid curves) for $p_1$, $p_2$, $q_2$ and $\psi$. Red points on the horizontal axis indicate the true parameter values.

Figure 6 shows that concentration of the posterior densities near the true parameter values, indicated by red points on the horizontal axis, is evident for $p_1$ and $p_2$ but less so for $q_2$ because of the smaller number of red vertices. We observed that posterior inference for these nuisance parameters were quite insensitive to the choice of hyperprior for $\psi$, even though the posterior distribution for $\psi$ itself obviously did depend on the hyperprior.

To quantify nomination performance, we repeated the simulation for 1000 graphs obtained using the same parameter settings, i.e. $n = 12$, $m = 5$, $m' = 2$, $p_1 = 0.25$, $p_2 = 0.15$ and $q_2 = 0.25$. The estimated probability of correct nomination based on these 1000 graphs is 0.44, with equal-tail 95% confidence interval estimated by the BCA bootstrap (Efron, 1987) as (0.41, 0.47). Recall that for this toy example, the probability of correct nomination purely by chance is 0.3. Hence, we will do significantly better than chance. The odds ratio for correct nomination relative to chance is $(0.44/0.56)/(0.3/0.7) = 1.8$.

Since we have, for any given graph, the marginal posterior probability that the nominated vertex is red, we can estimate the conditional probability of correct nomination given that this marginal posterior probability exceeds *p*. This is shown in Figure 7(a) where, for example, if the marginal posterior probability that the nominated vertex is red exceeds 0.4, then the conditional probability of correct nomination is estimated to be 0.55, with 95% confidence interval of (0.49, 0.60). There is evidence of a trend of increasing probability of correct nomination with increasing posterior probability that the nominated vertex is red.

The marginal distributions of the posterior means obtained from the 1000 graphs, for the nuisance parameters and hyperparameter, are illustrated in Figure 8(a). They show concentration of probability mass around the true values of the nuisance parameters.

Figures 2 and 4 suggest that a smaller number of MCMC iterations may be required for inference about posterior means, and hence about the probability of correct nomination. To check whether this is the case, we looked at results using 1000 MCMC iterations after a burn-in of 1000 iterations. The estimated probability of correct nomination for the same 1000 graphs was 0.44, with equal-tail 95% BCA bootstrap confidence interval of (0.41, 0.46), which is almost identical to what we had before with 10 times the number of iterations. Plots corresponding to Figures 7(a) and 8(a) for the reduced number of iterations are given in Figures 7(b) and 8(b), respectively. The similarities are evident, indicating little change with the reduced number of iterations. Thus, considerable computational savings can be achieved by restricting inference to posterior means and using the smaller number of iterations. This holds also for the experiments with the Enron data in the next section, hence the use of the smaller number of iterations there.

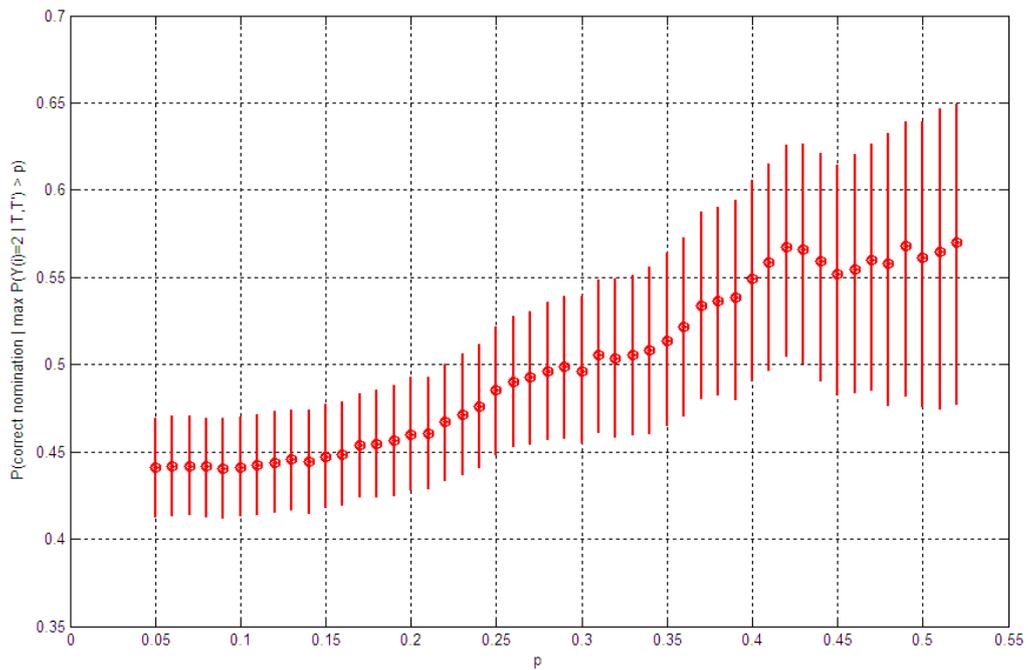

(a)

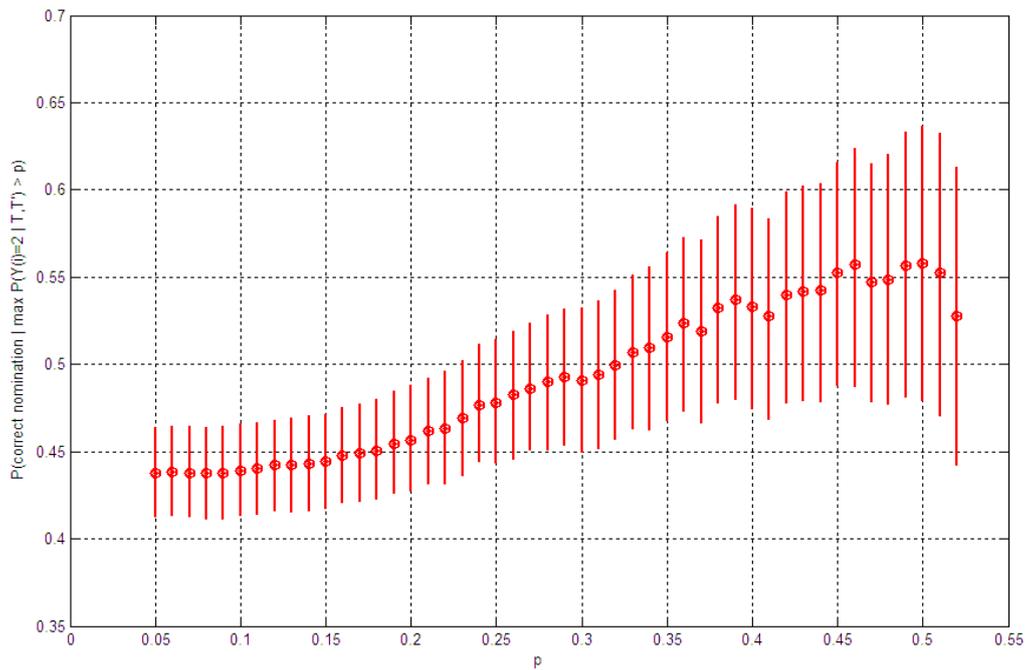

(b)

Figure 7. Conditional probability of correct nomination given that the marginal posterior probability that the nominated vertex is red exceeds $p$, with equal-tail 95% BCA bootstrap confidence intervals. This is obtained from 1000 graphs, each with (a) 10000 MCMC iterations after a burn-in of 10000 iterations; (b) 1000 MCMC iterations after a burn-in of 1000 iterations.

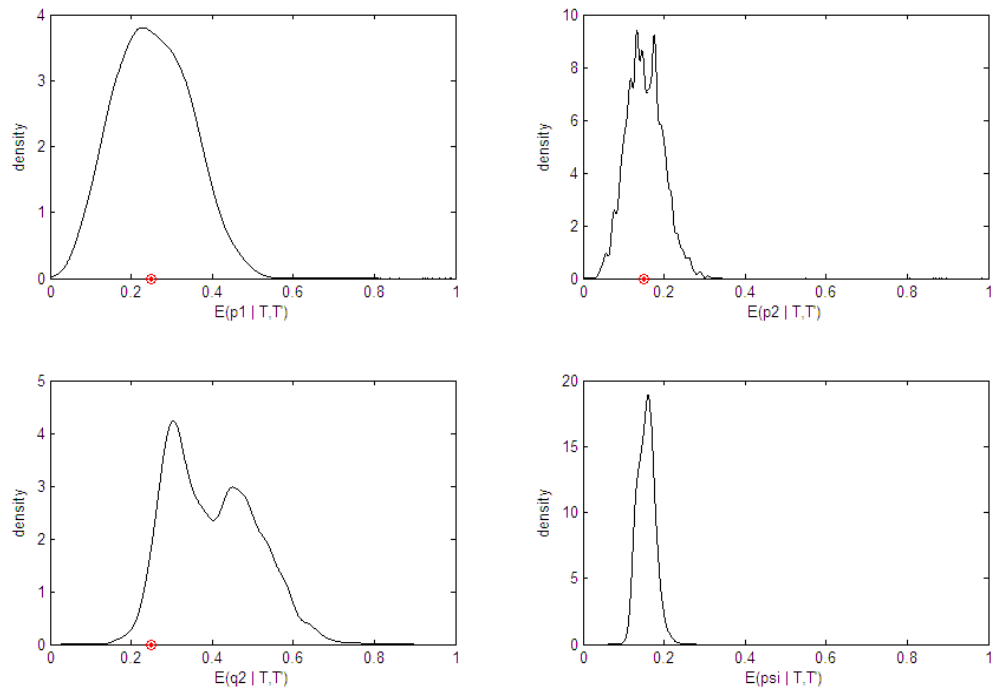

(a)

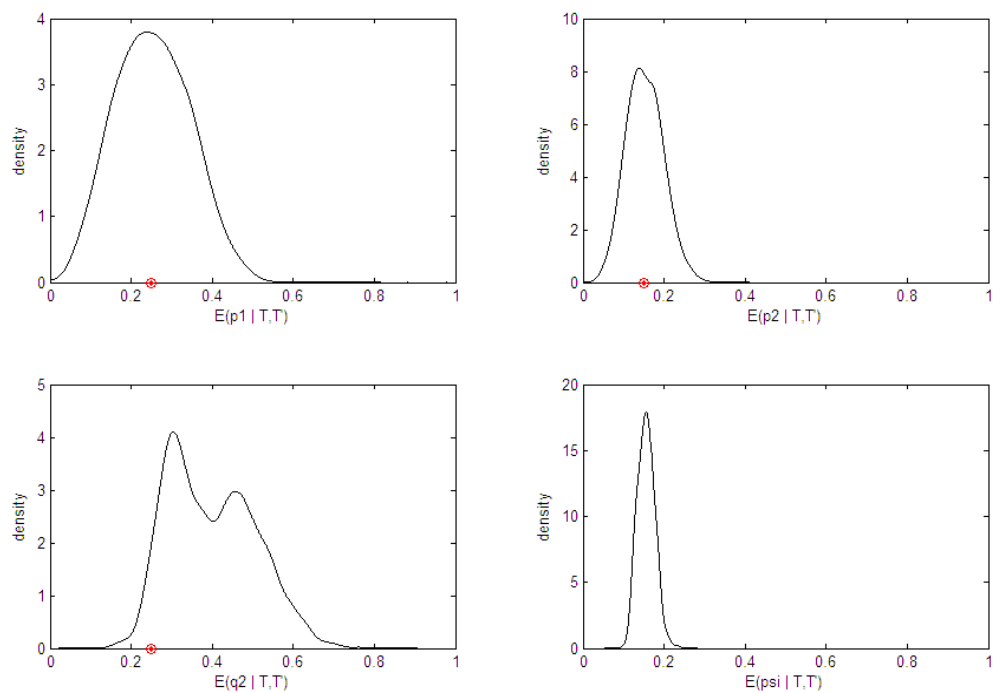

(b)

Figure 8. Kernel densities fitted to posterior means for $p_1$, $p_2$, $q_2$ and $\psi$, obtained from 1000 graphs. Red points on the horizontal axis indicate the true parameter values. This is obtained using (a) 10000 MCMC iterations after a burn-in of 10000 iterations; (b) 1000 MCMC iterations after a burn-in of 1000 iterations.

Coppersmith and Priebe (2012) defined a linear fusion statistic for vertex *v*, combining its context and content statistics, as

$$\tau_\lambda(v) = (1-\lambda)R(v) + \lambda S(v), \tag{37}$$

where $\lambda \in [0, 1]$ is a fusion parameter that determined the relative weight of context and content information. For a given value of $\lambda$, the nominated vertex was a latent vertex with the largest value of $\tau_\lambda$. To compare that approach with the one here, we conducted a simulation study adopting the values that they had used: $n = 184$, $p_1 = 0.2$, $p_2 = 0.2$ and $q_2 = 0.4$. We investigated two values for *m*, 8 and 32, which represented a "small" value and a "large" value of *m* that they had considered. For each value of *m*, we looked at $m' = m/4$, $m/2$ and $3m/4$, just as they had done. The results, in terms of the probability of correct nomination estimated from 1000 graphs, are given in Tables 3(a) and (b).

Table 3. Estimated probability of correct nomination based on 1000 graphs, for $n = 184$, $p_1 = 0.2$, $p_2 = 0.2$ and $q_2 = 0.4$. BVN denotes the approach in this paper while C&P is the approach described in Coppersmith and Priebe (2012).

(a) $m = 8$

|  | $m' = 2$ | $m' = 4$ | $m' = 6$ |
|---|---|---|---|
| BVN | 0.09 (0.08, 0.10)[†] | 0.12 (0.10, 0.13) | 0.09 (0.08, 0.11) |
| C&P[††] | 0.09 | 0.11 | 0.06 |
| OR($\frac{BVN}{C\&P}$)[†††] | 1 | 1.10 | 1.55 |

(b) $m = 32$

|  | $m' = 8$ | $m' = 16$ | $m' = 24$ |
|---|---|---|---|
| BVN | 0.83 (0.81, 0.85) | 0.90 (0.88, 0.92) | 0.87 (0.85, 0.89) |
| C&P | 0.83 | 0.86 | 0.78 |
| OR($\frac{BVN}{C\&P}$) | 1 | 1.47 | 1.89 |

[†] 95% BCA bootstrap confidence interval.
[††] Optimal performance with optimal fusion parameter.
[†††] Odds ratio for correct nomination.

In the tables, BVN denotes the approach in this paper, for which the estimated probability of correct nomination and 95% BCA bootstrap confidence interval are given. C&P denotes the approach in Coppersmith and Priebe (2012), for which the estimated probabilities of correct nomination that are shown corresponded to optimal values of the fusion parameter that gave the best performance. The last row in each table gives the odds ratio (BVN relative to C&P) for correct nomination. We see, for this simulation study and in terms of the probability of correct nomination, that the two approaches have the same performance when $m' = m/4$, i.e. when the number of latent red vertices, $m - m'$, is big relative to $m$. However, as the number of latent red vertices gets smaller relative to $m$, BVN performs increasingly better than C&P (for $m = 8$, the odds ratio increasing from 1 to 1.10 to 1.55 as $m'$ increases and, for $m = 32$, increasing from 1 to 1.47 to 1.89). Moreover, the rate of improvement appears to be higher for a bigger $m$. Note that in an actual application, the C&P approach might not achieve its optimal performance because of the difficulty of finding the optimal value of the fusion parameter.

## 5 Application Results

The Enron email corpus, available at http://www.enron-mail.com/, consists of email communications amongst Enron employees and their associates. Some of them were allegedly committing fraud and their fraudulent activity was captured in some emails along with many innocuous ones. Priebe, et al. (2005) derived a processed version of a subset of the email data, over a period of 189 weeks from 1998 to 2002. This yielded 189 graphs (1 graph per week), each containing the same 184 email users forming the vertices of the graph. 10 of these users have been found to have committed fraud.

Berry, et al. (2007) indexed the contents of a subset of the email corpus into 32 topics. These same topics were adopted by Coppersmith and Priebe (2012), who introduced a mapping from the topics to a binary edge attribute, {green, red}, denoting content perceived as innocuous and fraudulent, respectively. We used one of the graphs derived by Priebe, et al. (2005), together with the binary edge attributes from Coppersmith and Priebe (2012), for the experiments described here.

For the first experiment, we used Priebe, et al.'s Enron graph for week 38. We treated 5 of the 10 fraudsters as observed red vertices, the other 5 as latent red vertices and all other remaining users as latent green vertices, to see whether one of the latent red vertices will be correctly nominated. The probability of correct nomination was estimated from all 252 (10 choose 5) combinations of 5 observed red vertices taken from the 10 fraudsters. For each combination, 1000 MCMC iterations were used for the estimation after a burn-in of 1000 iterations. The estimated probability of correct nomination is 0.10, with 95% BCA bootstrap confidence interval of (0.09, 0.11). Note that for this experiment, the probability of correct nomination purely by chance is $5/179 \approx 0.03$. This gives an odds ratio of $(0.1/0.9)/(0.03/0.97) = 3.6$, for correct nomination by the Bayesian model relative to chance.

The distributions of the marginal posterior means of the nuisance parameters, from the 252 combinations, are shown in Figure 9. The sample means of the posterior means are $\bar{p}_1 = 0.0168$, $\bar{p}_2 = 0.0111$ and $\bar{q}_2 = 0.1298$. These estimates will be used in the second experiment.

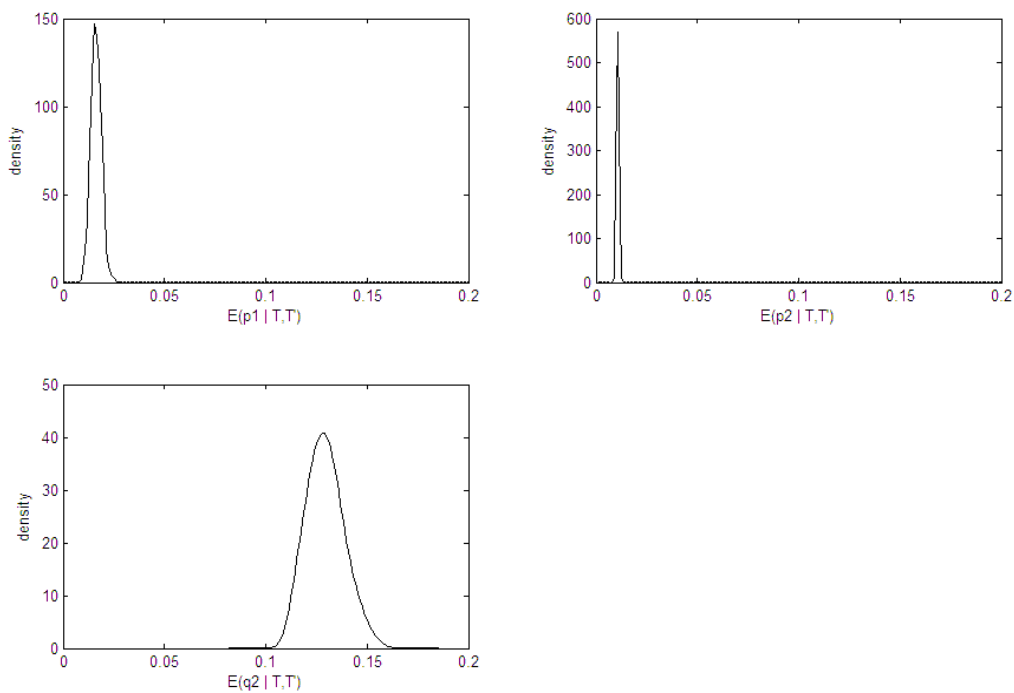

Figure 9. Kernel densities fitted to posterior means for $p_1$, $p_2$ and $q_2$, obtained from the 252 combinations in Experiment 1.

For the second experiment, we treated the estimates of $p_1$, $p_2$ and $q_2$ from the first experiment as true values in a Monte Carlo simulation involving $n = 184$, $m = 10$ and $m' = 5$. Vertices 1 to 5 were known red vertices, 6 to 10 were latent red vertices and the rest were latent green vertices. 1000 graphs were generated and processed, each with 2000 MCMC iterations. The first 1000 iterations were discarded and the remaining 1000 iterations used for inference. The probability of correct nomination was estimated to be 0.50, with 95% BCA bootstrap confidence interval of (0.47, 0.53). Hence, the odds ratio for correct nomination relative to chance is $(0.50/0.50)/(0.03/0.97) = 32.3$.

Estimates of the probability of correct nomination given that the posterior probability that the nominated vertex is red exceeds $p$ are illustrated in Figure 10. Once again, there is a clear trend of increasing probability of correct nomination with increasing posterior probability. Furthermore, since we have more data (larger graph with more vertices), the probability of correct nomination is higher. For example, recall that for the graph with 12 vertices in the previous section, the probability of correct nomination given that the posterior probability exceeds 0.4 was between 0.49 and 0.60 with 95% confidence. For the current graph with 184 vertices, the same probability of correct nomination is between 0.67 and 0.78.

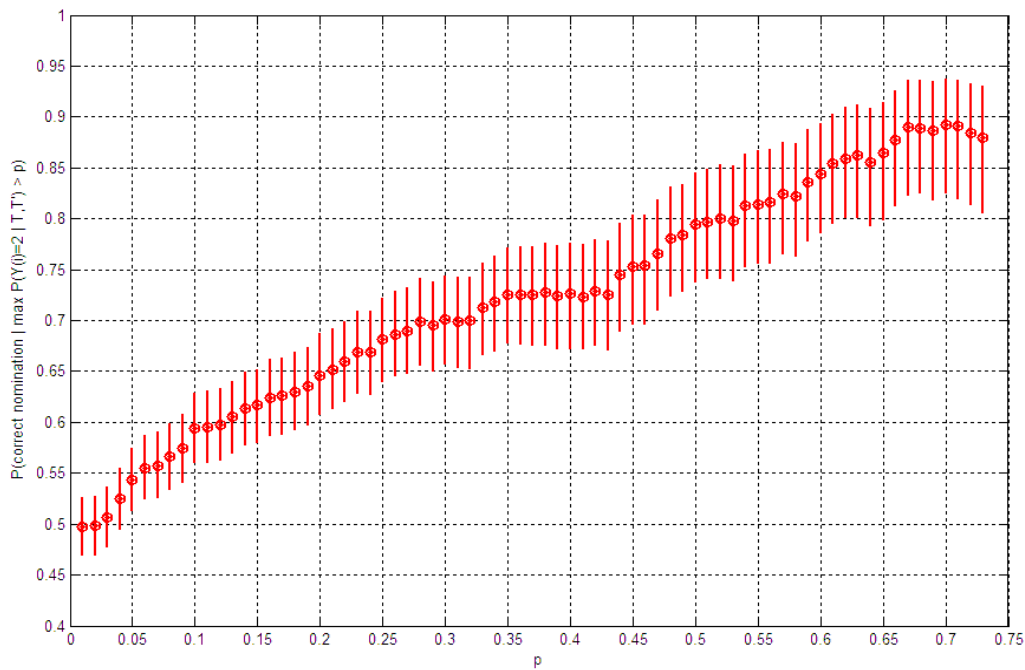

Figure 10. Conditional probability of correct nomination given that the marginal posterior probability that the nominated vertex is red exceeds $p$, with equal-tail 95% BCA bootstrap confidence intervals. This is obtained from 1000 graphs, each with 1000 MCMC iterations after a burn-in of 1000 iterations.

## 6 Conclusion

We have formulated a Bayesian model for the vertex nomination problem, based on the likelihood model proposed by Coppersmith and Priebe (2012). Inference with the model proceeds via a Metropolis-within-Gibbs algorithm for generating sample points from the posterior distribution. Results from a simulation study show that the Bayesian model performs significantly better than chance. Furthermore, there is evidence of a trend of increasing probability of correct nomination with increasing posterior probability that the nominated vertex is red. Similar results are observed from experiments with the Enron email corpus. Another simulation study, comparing with the approach in Coppersmith and Priebe (2012), shows that the Bayesian model performs increasingly better as the number of latent red vertices gets smaller relative to the total number of red vertices, and with a higher rate of improvement when the total number of red vertices is bigger.

## Acknowledgements

Thanks to Youngser Park for assisting with the Enron data and for helpful comments on a draft version. Thanks also to Celine Cattoen-Gilbert for suggestions for speeding up computations.